\documentclass[journal,twoside]{IEEEtran}
\usepackage{graphicx}

\ifCLASSINFOpdf
\else
\fi
\begin{document}
%
\title{Stimulating Higher Order Thinking in Mechatronics by Comparing PID and Fuzzy Control}
%
%
%
\author{
\begin{tabular}[t]{c@{\extracolsep{8em}}c} 
Christopher J. Lowrance  & John R. Rogers \\
Artificial Intelligence Task Force & Department of Civil and Mechanical Engineering \\ 
US Army Futures Command & United States Military Academy \\
Email: christopher.j.lowrance.mil@mail.mil & West Point, NY USA\\
& Email: john.rogers@westpoint.edu
\end{tabular}
}
\markboth{Computers in Education Journal, Volume 10, Issue 3, September 2019}%
{Lowrance and Rogers: Stimulating Higher Order Thinking in Mechatronics by Comparing PID and Fuzzy Control}



\maketitle

\begin{abstract}
Many studies have found active learning, either in the form of in-class exercises or projects, to be superior to traditional lectures.  However, these forms of hands-on learning do not always get students to reach the higher order thinking skills associated with the highest levels of Bloom's Taxonomy (i.e., analysis, synthesis, and evaluation). Assignments that expect students to take an expected approach to reach a well-defined solution contribute to a lack of higher order thinking at the college level. Professional engineers often face complex and ambiguous problems that require design decisions, where there is no straightforward answer. To strengthen the higher order thinking skills that these problems demand, we developed a project in our semester-long mechatronics course where students must evaluate two automatic control methodologies for an application without being given explicit performance criteria or experimental procedures.  More specifically, the project involves determining the superior control method for leader-follower behavior in which a ground vehicle autonomously follows behind a lead vehicle. Laboratory exercises throughout the semester expose the students to the skills they need for the project: using sensors and actuators,  programming a proportional-integral-derivative (PID) controller and a fuzzy controller, and using computer vision to detect the signature of an object.  In the final course project, they go beyond implementing individual controllers and create their own evaluation criteria and experiments for making a design decision between PID and fuzzy control.  We implemented this approach for three semesters, and our significant findings are:  1) students generally appreciate the aspect of working on a real-world and open-ended problem, 2) most teams developed creative performance criteria and methods for evaluating controller performance, clearly demonstrating higher order thinking, and 3) students discover that creating  a comparative study is nontrivial due to the number of factors that influence performance, which mimics the practical problems they will likely face as engineers.   
\end{abstract}


%
\IEEEpeerreviewmaketitle

\section{Introduction}

Generally, real-world problems are not as straightforward as the homework assignments and lab exercises assigned to students in traditional engineering courses. In other words, engineers in the workplace will not be given a set of instructions and will not merely have to recall and apply a formula to solve a problem, yet many engineering curricula still do not have opportunities for students to develop their critical thinking and problem-solving skills.  As a result, students are sometimes unable to transfer concepts learned in the classroom to real situations when presented with open-ended design problems \cite{lane}.

 \begin{figure}
\centering
\includegraphics[width=3.5in]{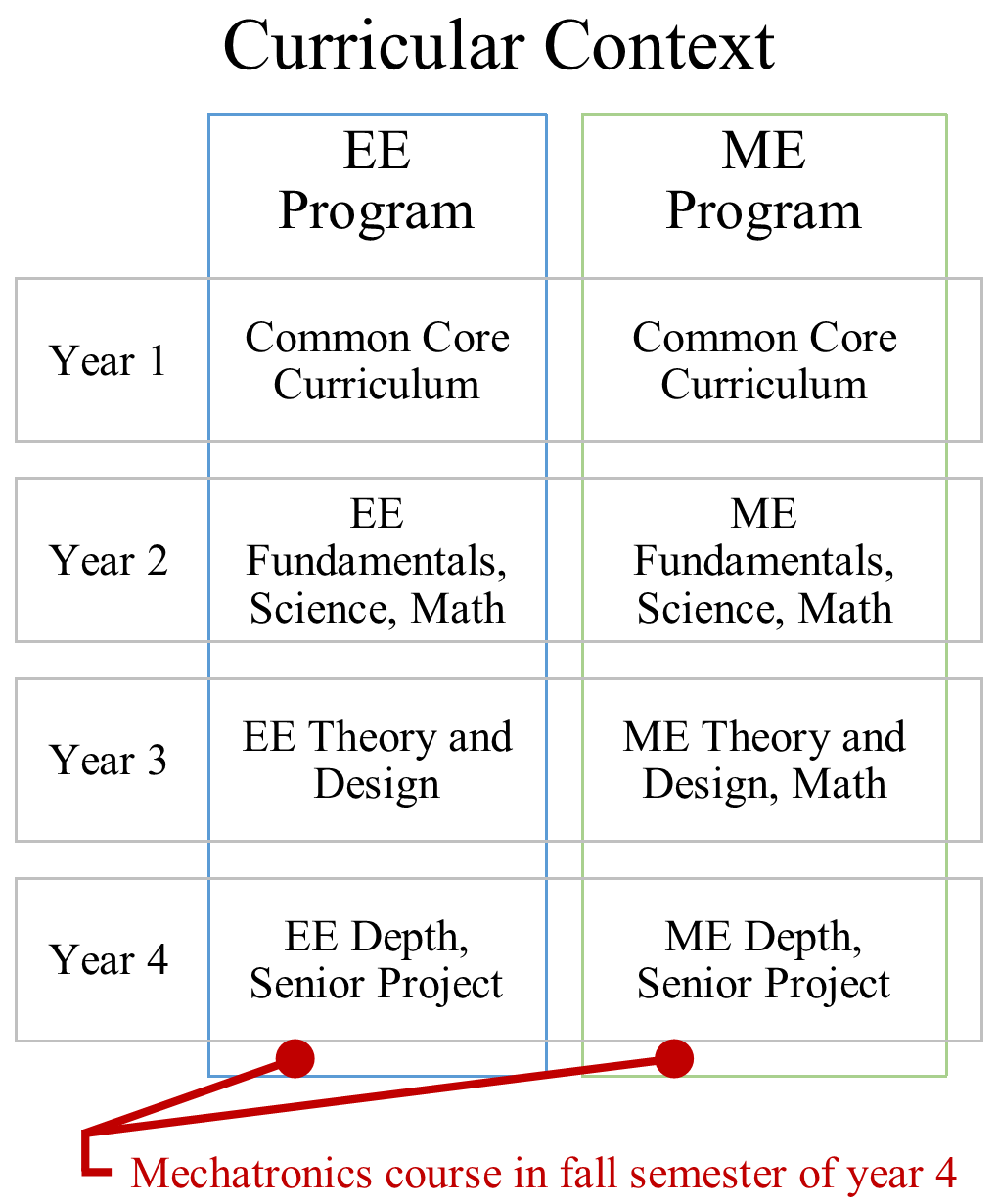}
\caption{The Mechatronics course is typically taken in the first semester of the senior year. It is also part of the Robotics minor. Almost all students that enroll are electrical engineering or mechanical engineering majors.}
\label{fig-1}
\end{figure}
 
 Engineers must be able to solve open-ended problems and make informed design decisions.  As part of that process, they need to be able to fairly evaluate alternatives and choose the best option among multiple possible solutions.  These kinds of real-world engineering design problems require practitioners to think critically, develop their own evaluation criteria, and construct personal knowledge based on experiments that they design. These tasks require higher order thinking skills that associate with the higher levels of Bloom's Taxonomy: analysis, synthesis, and evaluation \cite{bloom}. Unfortunately, many schools do not teach this type of engineering analysis and decision making until their senior design course.  

Project-based learning is a form of active learning that can promote higher order thinking \cite{savery}.  Over the past several years, we have employed project-based learning in our undergraduate mechatronics course \cite{rogers}.  As part of our hands-on approach, we used a series of mini-projects to build competencies for an end-of-course autonomous robot project.  

In prior semesters, the projects focused merely on building mechatronic systems without necessarily analyzing design alternatives and evaluating performance.  Although some of the projects were open-ended and allowed students to solve them in different ways, we did not require them to consider the alternative options or defend their design choices based on performance analysis and open debate.  Instead, grades were based on whether their end project functioned properly and met some set of basic specifications.  In other words, there was no requirement for them to think deeply about their design, quantify its performance using their own set of measurable metrics, and then optimize the design.  

This lack of analysis led us to improve upon our project-based course and make a final project that requires students to evaluate design alternatives more thoroughly and quantify performance differences.  In the process of conducting such an experiment, our students are required to think more deeply, and in the process, reach a higher order of thinking described by the verbs in upper three levels of Bloom's taxonomy (i.e., analysis, synthesis, and evaluation).  Whereas in the past, some of our projects may have only required the students demonstrate a level of thought equivalent to the next lower two levels (i.e., comprehension and application).   

The purpose of this paper is to detail our final project that is intentionally structured with the aim of strengthening the higher order thinking skills of our graduates.  In this paper, we expand upon an earlier version \cite{original} by adding new figures to aid in the explanation of the project and strengthening the analysis based on the feedback captured from another semester of teaching the course.  The culminating project we used for the last three years is a comparative study of two different control methods implemented earlier in the course.  The application scenario given to the students is to determine whether PID or fuzzy control is optimal for automating the drive system of a mobile robot that must follow a lead vehicle in compliance with a set of specifications.  This type of autonomous formation control is commonly referred to as \textit{leader-follower} \cite{mariottini}.    
 
Mechatronics is a multidisciplinary field that crosses the traditional boundaries of mechanical, electrical, and computer engineering, and loosely defined as the applications of those fields to the design of products or systems \cite{shooter}.  At the United States Military Academy (USMA), mechatronics is a senior-level undergraduate course.  It is an interdisciplinary and hands-on class taken mostly by mechanical and electrical engineering students and taught by a team of professors from each of those disciplines.  Students from other majors, such as computer science, may elect to take the course if they meet the prerequisites, which include classes in circuit analysis and control theory.  At USMA, mechatronics is part of a sequence of electives that students may take as part of the robotics-depth option within the electrical engineering program, the mechatronics track within the mechanical engineering program, and the robotics minor within the academy curriculum, see Fig. \ref{fig-1}. Given the course's connection to a thread of other classes, it is intentionally structured with a controls and robotics theme.  Similar to mechatronics courses offered at other colleges, our course exposes the students to rapid prototyping and other topics associated with mechanical engineering, but it mostly focuses on hardware (sensor) integration and software coding (e.g., event-driven programming and discrete control) given that controls and sensors are embedded into most mechanical systems \cite{shooter}.     

\section{Related Work}
Higher order thinking skills such as critical thinking go beyond basic memorization and the application of facts, and they are the types of skills needed to be evaluative, creative, and innovative \cite{center}.  The well-known classification model used to quantify educational learning objectives concerning cognitive complexity referred to as Bloom's taxonomy is also associated with a hierarchy of higher order thinking skills \cite{lane}.  The level of cognitive processing required by these thinking skills increases as one progresses up the hierarchy.  Educators at the college level usually use Bloom's taxonomy as an aid when structuring and assessing the learning objectives of their courses, and generally, they want portions of their courses to require students to utilize and strengthen skills at the higher layers of the hierarchy \cite{center}.  Higher order thinking skills are typically associated with the top three levels of Bloom's taxonomy (analysis, synthesis, and evaluation) and include skills such as critical thinking and problem-solving \cite{asok}.  These skills are engaged when students face the uncertainty of design questions and open-ended problems \cite{asok}.  
With the goal of developing higher-order thinking, educators have used a mix of strategies in engineering classes.  Asok et al. presented some ideas, such as role-playing and debating, which could be used as in-class exercises \cite{asok}.  Similarly, Lane and Farris had students openly discuss their solutions to engineering problems as in-class exercises, and the class was expected to think critically about the proposals and offer feedback \cite{lane}.  Bee converted an engineering course from lecture-based over to a problem-based learning format \cite{bee}. By doing so, Bee discovered that students were more motivated and appreciative of working on real-world problems; consequently, students tended to research problems to greater depth \cite{bee}.  Madhuri et al. adopted an inquiry-based learning model so that chemistry experiments were contextually related to real life; furthermore, students had to openly answer questions and discuss concepts in pre-lab sessions before conducting their experiments \cite{madhuri}.  Pinho-Lopes and Macedo found project-based learning to be effective in promoting higher-order thinking in two civil engineering courses and used student performance along with questionnaires as means of assessment \cite{pinho}.  

Mechatronics generally lends itself to collaborative and active learning strategies such as project-based learning \cite{shooter,mynderse}.  Additionally, the interdisciplinary subject provides an excellent opportunity to build real-world problem-solving skills because it requires students to work outside their discipline as practicing engineers must often do. \cite{deanes}.  From the literature, it appears that most mechatronic courses take a graduated approach to project-based learning where initial lab exercises reinforce fundamentals that are then applied in a more challenging final project, and the general intent of the final project is usually to strengthen the student's design and problem-solving skills.  For instance, Mynderse and Shelton use more traditional lectures and lab experiences at the beginning of the course and then transition to designing and building an autonomous vehicle as a culminating project \cite{mynderse}.  Gurocak employs small-scale projects at the beginning of the course to build competencies for the later culminating class project \cite{gurocak}.  Similarly, Shooter and McNeill use a combination of active learning methods including in-class exercises, team homework, and lab assignments at the beginning and then devote the final five weeks to a design project \cite{shooter}.  Consi emphasizes creativity in a mechatronics project by assigning students to ``create something" using a stepper motor-driven XYZ platform as is used in 3D printers and CNC milling machines \cite{consi}.

Many of the design projects in these courses are open-ended, allowing the students the creative freedom to design a solution.  However, it is unclear whether the students are genuinely making design choices by weighing alternatives in an engineering manner, or whether they are building a solution based on using the first option that works.  As a step in the right direction, some courses require students to explain their designs to their peers who act as critics alongside the professor \cite{lane,mynderse,gurocak}.  But overall, there seems to be a lack of deeply analyzing design choices, as well as optimizing and evaluating a particular design selection.  

Many of the skills associated with higher-order thinking overlap with those that are essential in self-directed and lifelong learning.  Self-directed learning (SDL) relates to a person's openness and effectiveness of learning, as well as their ability to conduct research and solve problems \cite{jiusto}. Furthermore, SDL encompasses a set of skills that include the ability to retrieve information, think critically, and efficiently communicate \cite{jiusto}.   

Courses that employ SDL involve a high degree of student autonomy and research where they must decide what is to be learned, determine an approach to learning, and manage the learning process independently \cite{litzinger}.  Some examples of engineering courses at the undergraduate level that rely heavily on SDL are capstones (senior design) and independent studies \cite{litzinger}, both of which are typically offered in many college curriculums in the United States.  However, other courses can embed elements of SDL, and these courses can develop SDL skills in learning environments that are slightly more structured such as project-based learning (PBL).  For example, Ulseth found extensive quantitative evidence that project-based learning graduates developed SDL abilities because the projects required the students to apply their knowledge, be reflective, and work under minimal supervision \cite{ulseth}.  A PBL approach taken by Jiusto and DiBiasio in their interdisciplinary global studies program, which includes teams of engineering and science majors working together on real-world problems, was shown to improve the SDL and lifelong learning skills of their university's graduates \cite{jiusto}. Fellows et al. develop the SDL skills and attitudes of their students early in their college experience through an interdisciplinary course intended for freshman \cite{fellows}; the course uses a series of projects that emphasize teamwork, research, problem-solving, and communications.

\section{Details about the Controller Comparative Study}
\subsection{Project Scenario and Setup}

\begin{figure}
\centering
\includegraphics[width=3.5in]{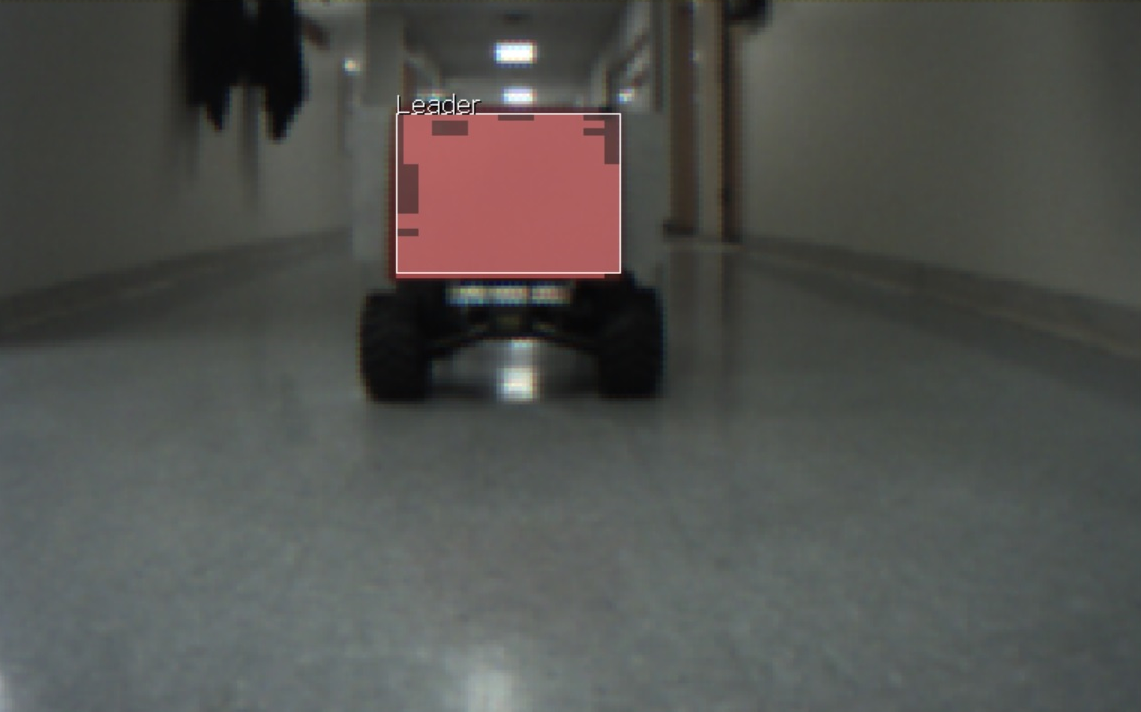}
\caption{Snapshot taken from the camera's software known as Pixymon. Pixymon can be used to configure the camera to recognize different colored objects and identify them uniquely.  Once the camera is configured to recognize a particular color, it automatically bounds that color with a box like the one shown in the figure.  The position and size of the box represents the pixel-level information that is sent to the microcontroller for control (tracking) decisions.}
\label{PixyView}
\end{figure}

For the final project in our course, students were given the task to experimentally determine whether a PID or fuzzy controller works best for automating the drive system of an autonomous ``follower" robot.  However, during the most recent iteration of the course, we also gave the students the option of proposing their own final project, but we stipulated that it must involve comparing the two control methodologies (PID and fuzzy) in the context of some relevant application.  Our intent was to incentivize the students and provide them the creative freedom to work on an application of their interest.  Nevertheless, only two out of the 11 groups proposed projects that met the guidance, while the others elected to compare the controllers in the context of leader-follower.  

In the project handout, we provided students with a specification but not an explicit set of instructions on the kinds of steps that they should perform to complete such a comparative study, and we provided no details on how they can conclude which is optimal. The students also conduct a literature review that may include conflicting reports of which controller type (i.e., fuzzy or PID) performs best under similar conditions.  We intended to keep the implementation of the controllers and the empirical analysis open-ended by not providing the students with a step-by-step procedure and not steering them onto a pre-approved solution. Thus, students are deliberately put in a position where creativity, independent ideas, and critical thought are required to determine the best approach.  We framed the project in a real-world context emulating practicing engineers faced with design decisions to meet specifications, who at times find conflicting reports as to which solution may be best for a particular problem.  Sometimes engineers must conduct their own independent and self-directed experiments to find the optimal solution.

Students worked on the project in teams of two unless there was an odd-numbered enrollment, in which case one group of three was formed.  The small team size encouraged participation and involvement from both members.  The instructors assigned the teams to make them interdisciplinary.  In most cases, they consisted of one electrical engineer and one mechanical engineer. 

The mobile robots used in the project were Traxxas E-Maxx vehicles which were modified so that they responded to pulse width modulation (PWM) control signaling from a microcontroller.  Students developed discrete control algorithms without the aid of software toolboxes or libraries. The control algorithms automated the steering and speed of the vehicle based on feedback provided by a camera configured to detect and bound the color of a ``leader" vehicle.  The camera used in the project was a Pixy \cite{pixy}, which uses onboard image processing to track colored objects.  An example of the Pixy's ability to recognize and bound colored objects within the context of an image is shown in Fig. \ref{PixyView}.  The image was taken from the Pixy of a follower robot that was configured to recognize red-colored paper affixed to a leader robot. The Pixy was programmed to tag the bounding box of the largest color-matched target with the word `Leader'--visible in Fig. \ref{PixyView}.

For every picture frame, the Pixy serially transmitted to the microcontroller the relative size and position of any colored object that it was configured to detect. The transmitted information included:  \textit{x} and \textit{y} pixel coordinates that corresponded to the center of the bounding box around the colored object, the height of the box in number of pixels, the width of the box in number of pixels, and the area of the box in pixels squared.  In general, most students designed their steering controller based on the error in the leader's x-axis position from the center of the frame, and the speed controller using the area (i.e., size) of the leader's signature and its difference from the desired set point.      

\begin{figure}[!t]
\centering
\includegraphics[width=3.5in]{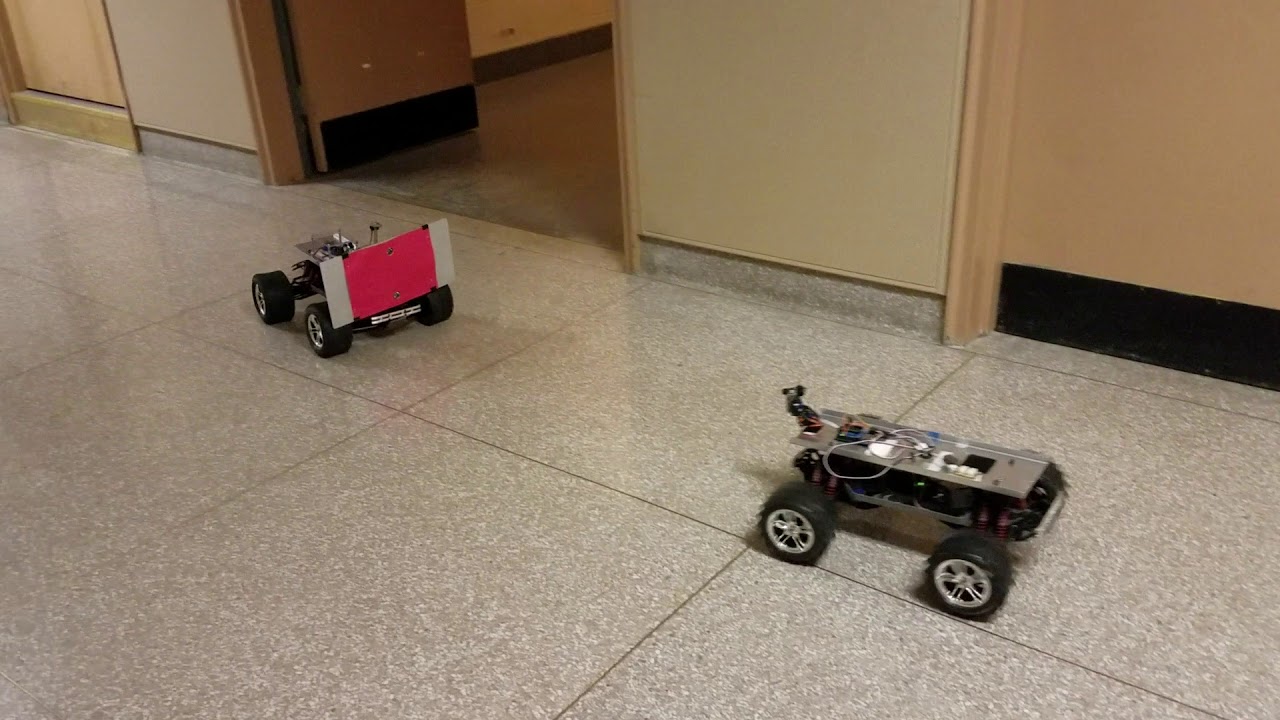}
\caption{Picture showing a pair of leader-follower robots in the hallway of our academic building. The camera mounted on the front of the follower vehicle is visible. The red paper on the leader is an easy-to-find target for computer vision.}
\label{botsHallway}
\end{figure}

To configure the leader with a trackable object for the Pixy, the rear of the leader vehicle was mounted with a flat panel that had a single colored 8.5" x 11" paper attached to it.  The flat, colored surface was secured perpendicular to the ground so that it faced the follower robot.  The camera on the follower was fixed to point directly to its front.  Fig. \ref{botsHallway} shows the configuration of the robots.  

Videos are posted on the West Point Robotics YouTube channel demonstrating the operability of the controllers developed by the students as part of the project.  The videos are organized in playlists, and the ones associated with this class are located under the course's title ``XE475" (see \cite{youtube}).  The leader vehicle in these videos was remotely controlled by a student while the trailing robot was programmed to autonomously follow the leader based on the speed and steering controllers developed by each team.

\subsection{Preliminary Exercises Leading to the Project}
At the beginning of the course, we teach the students the fundamentals of integrating sensors and creating event-driven (e.g., state machine) programs for mechatronic systems using microcontrollers.  The students use the Arduino microcontroller and its integrated development environment (IDE) in all course projects.  The Arduino IDE abstracts much of the register-level details required to program most other types of microcontrollers.  Thus, the high-level IDE allows the students to focus their time on developing and tuning mechatronic algorithms (e.g., control algorithms) rather than studying the fine-grained details of embedded programming that we cover in another course.  

Previous mini-projects also expose the students to the process of how to design, program, and tune PID and fuzzy controllers using the Arduino.  For these projects, the students are required to develop their own Arduino code based on the C programming language without using any prewritten library for PID or fuzzy control.  By not relying on an abstracted function for these controllers, it ensures the students have a deeper understanding of the underlying principles of each control method, and it affords them greater flexibility for controller optimization during the comparative study.  Before the final project, one exercise involves implementing a PID position controller for a motor using feedback from a magnetometer mounted on a shaft.  Subsequently, the next two mini-projects have the students develop PID and fuzzy controllers for the robot control application of leader-follower.  By assigning these individual mini-projects before the start of the comparative study, it familiarizes the students with each controller type and also makes the final project more manageable so that students can complete it in the remaining part of the semester.   

The final project is a quantitative study that compares the controllers in a data-driven fashion using controller feedback to precisely quantify the follower's ability to track the path of the leader robot.  Students prepare for this type of in-depth analysis by completing preliminary projects that require them to use the data logger peripheral for the Arduino.  The course emphasizes the use of data logging in multiple projects as a means to characterize system performance beyond visual inspection.  In these preliminary exercises, the students are expected to record data and determine whether their systems meet a project's specifications and to include such evidence, in the form of statistics or visualizations, in their project reports.  The final project requires the same type of data analysis in the comparative study. 

\subsection{In-Progress Reviews for Tracking Progress, Exchanging Ideas, and Motivating Participation}
Student progress on the project was monitored through a series of in-progress reviews (IPRs).  One IPR was scheduled every week over the roughly four-week project.  Each status update had a high-level agenda as outlined in Table 1.  During the IPRs, the students briefed the entire class on their current findings. 

\begin{table}[h!]
  \begin{center}
    \caption{Project In-Progress Reviews}
    \label{tab:table1}
    \begin{tabular}{l|l} 
      \hline
      IPR 1 & Controller Optimization \& Tuning Results \\
      IPR 2 & Controller Comparative Figures (Preliminary)  \\
      IPR 3 & Controller Comparative Figures \& Statistical Results (Final) \\
      IPR 4 & Paper Outline \\
      \hline
    \end{tabular}
  \end{center}
\end{table}
The IPRs served multiple purposes.  First, they provided the students a high-level list of tasks to perform and a rough schedule to help guide them.  Secondly, the periodic IPRs were a way of formally checking whether teams were on track regarding the project timeline.  Lastly, the reviews facilitated early feedback to the students, so that if they needed it, a team would have sufficient time to make corrections to items, such as their experimental plans or data presentation methods.  

There were several reasons for having the teams present their IPRs openly to the class.  The student audience was encouraged to think critically about the approaches presented by their peers and to ask questions.  In the process, the briefings transformed the classroom away from an instructor-centered environment and more into an active learning environment where students could further develop their higher order thinking skills. Additionally, the presentations offered a way of sharing and exchanging ideas about the project with the entire class.  Also, we hypothesized that students would be more motivated knowing that they were subject to questions, not only from their professors but also from their peers.  Lastly, the briefings offered an opportunity for the students to sharpen their oral communication skills. 

\subsection{Final Reports in Scholarly Journal Format for Teaching Scientific Writing}

Several weeks before their final project, the students attended a lesson on the art of scientific writing and the composition of a scholarly engineering paper.  At that time, the students were briefed on the problem statement of their final project, and their homework was to review at least three research papers related to comparing fuzzy and PID control in various applications and write a synopsis on each.  As part of the same lesson, we also taught the students about the fundamentals of typesetting using \LaTeX and the benefit of vector graphics.  The motivation behind this exercise was twofold:  1) provide relevance to their project by showing them that others in the scientific community are studying such problems in other applications, 2) prepare them for writing quality reports and strengthening their technical writing abilities.

\section{Observations from the Project and Assessing Higher-Order Thinking}

\subsection{Example of Creativity \& Higher-order Thinking}
One of the primary objectives of the project was to get the students to think deeply about the problem and to develop their own set of experiments and criteria for analyzing controller performance.  Some typical examples of the students' approaches are illustrated in the following figures.  Fig. \ref{speedSetup} shows one of the usual strategies that teams took to expose the speed (i.e., throttle) controllers to step responses.  In a step response, the controller is supposed to abruptly transition from zero to maximum effort suddenly and then transition to steady-state once the error has been minimized, but determining how to expose the controller to such a test is not always straightforward.  For instance, the students had to ask themselves how far away they should place the follower robot from the stationary leader before activating the controller to correct the error.  In the process, most of them varied the start point of the follower robot in order to get the robot to exert maximum effort, and in the process, they discovered that the response was different depending upon the starting distance between the robots or how much error the follower needed to correct at the start of the test.  

\begin{figure}[!t]
\centering
\includegraphics[width=1.7in]{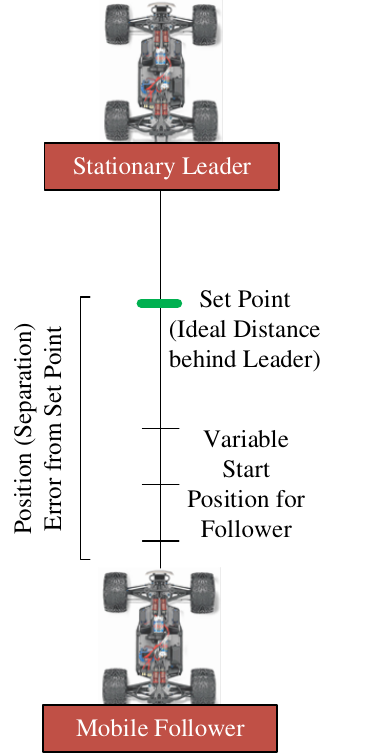}
\caption{Illustration showing the typical experimental setup developed by the students to subject the speed controller to different types of step responses.}
\label{speedSetup}
\end{figure}

\begin{figure}[!t]
\centering
\includegraphics[width=2in]{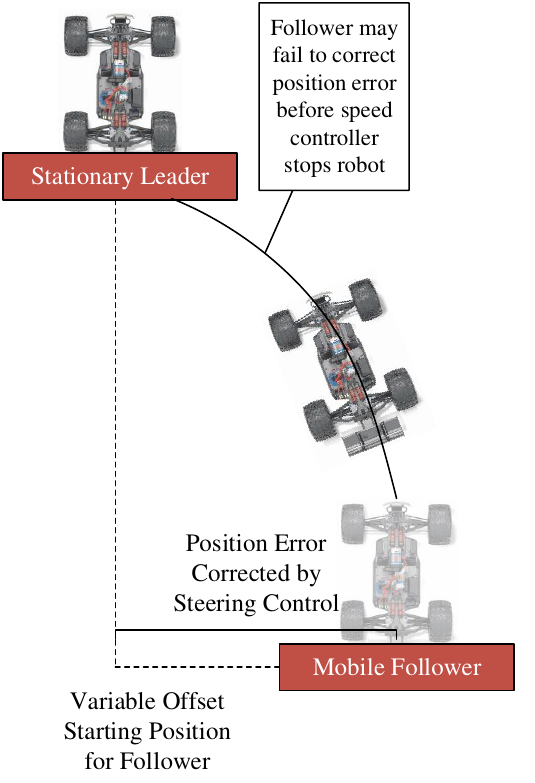}
\caption{Illustration showing one of the challenges encountered when some students attempted to evaluate their steering controllers.  They discovered that the controller either did not have the opportunity to correct its alignment or the follower ended up behind the leader, but at an angle that would eventually need counter correction.}
\label{steeringDifficulty}
\end{figure}

\begin{figure}[!t]
\centering
\includegraphics[width=2in]{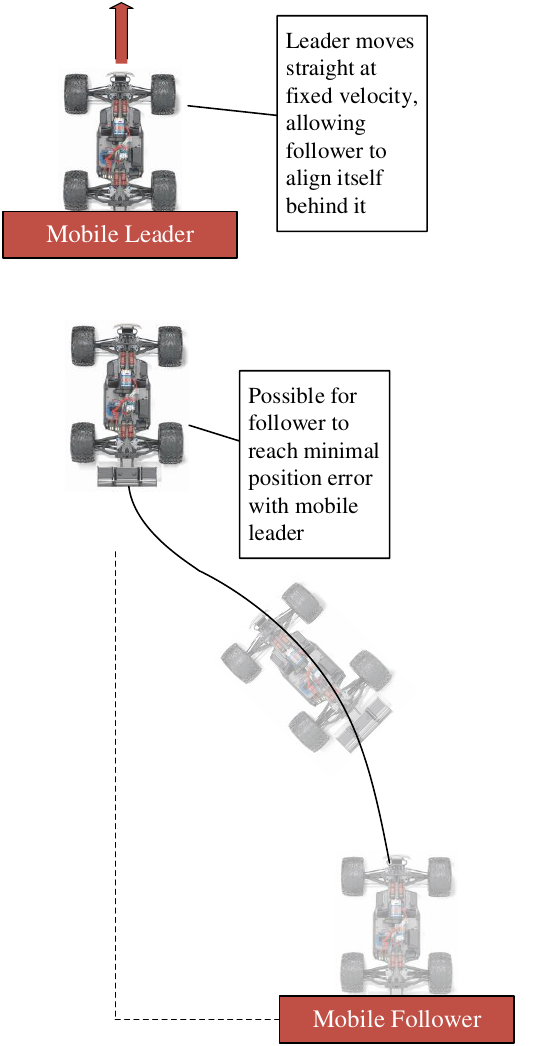}
\caption{Illustration showing how some teams overcame the challenge presented in Fig. \ref{steeringDifficulty} by having the leader move straight ahead at a fixed velocity.}
\label{steeringOvercome}
\end{figure}

The other controller that the students needed to evaluate was for the robot's steering.  However, as the students discovered, developing an experimental procedure for measuring the performance of the two controllers under these conditions was non-trivial.  Fig. \ref{steeringDifficulty} shows one of the difficulties encountered by some of the students.  Some students took the approach of merely offsetting the follower by some fixed distance and then expecting the controller to correct the alignment so that the follower moved in line with a stationary leader. They discovered that the follower could not minimize the error and reach steady-state before stopping due to the separation distance between the robots.  Other teams developed the idea of having the leader move forward at a fixed velocity throughout the test which eventually allowed the follower to align its wheels straight behind the leader. Fig. \ref{steeringOvercome} shows this concept.  Other challenging questions arose during the evaluation process, such as whether or not to engage the speed controller during the steering testing.  Some groups used a fixed velocity, and others varied the speeds to control the effects of the steering controller and test it at different rates of speed. 

Combined steering and throttle control could be a rich case study for analytical optimization rather than tuning (heuristic) optimization. There are two objectives from the point of view of the follower vehicle:  1) minimize the lateral distance from the track of the leader, and 2) maintain a desired following distance.  These objectives can be described mathematically with the proper coordinates and definitions. ``Maintain a desired following distance" for example is equivalent to ``minimize the deviation (position error) from the following distance (setpoint)."  The course does not currently treat the follower task as an analytical optimization problem, but it is an exciting and promising approach to consider for the future development of the course.  Students would benefit from a deeper understanding of the dynamic model of the system.  With an understanding of this model, the students could apply traditional optimization methods using analytical and simulation techniques.  For example, the vehicle kinematics could be simulated based on a no-slip bicycle model. In this model, the single rear wheel is always straight, and the single front wheel is steered on a vertical shaft through the point of contact. This model does not address understeer or oversteer, and it does not account for the complex three-dimensional dynamics of the vehicle, but it would be useful nonetheless as an introduction to computer-based modeling. This model can be used to predict the track of the vehicle for a given steering input (steering angle vs. time).

\begin{figure}[t]
\centering
\includegraphics[width=\linewidth]{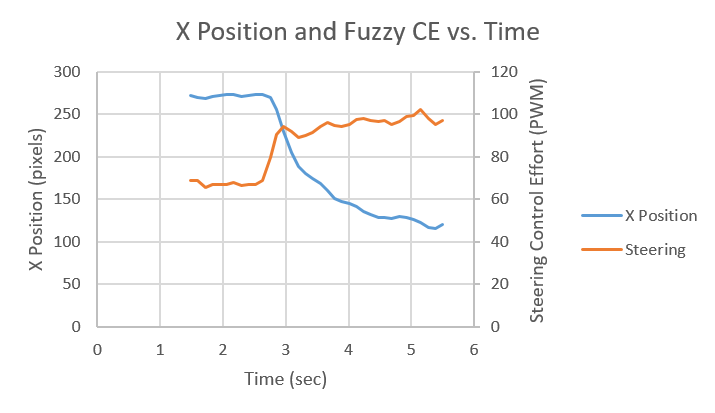}
\caption{Student results for fuzzy steering control.}
\label{figFuzzy}
\end{figure}

\begin{figure}[t]
\centering
\includegraphics[width=\linewidth]{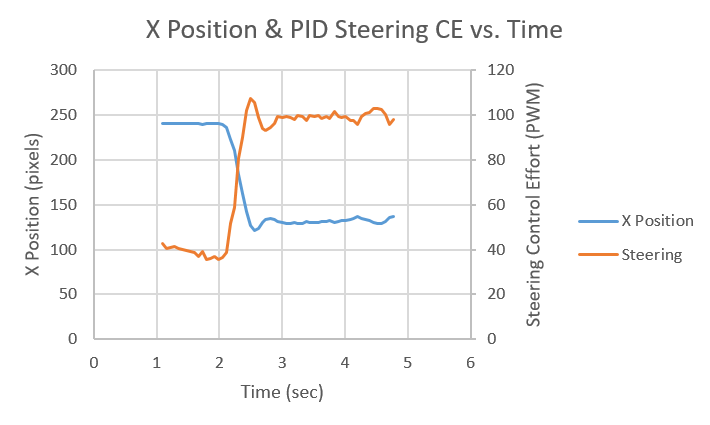}
\caption{Student results for PID steering control.}
\label{figPID}
\end{figure}

Figures \ref{figFuzzy} and \ref{figPID}  show the steering control effort and response for fuzzy and PID control; these are student results from their project report. The horizontal axis is time. This team used two vertical axes: the left axis, labeled “X Position (pixels),” is the location of the centroid of the detected target measured horizontally from the center, in pixel units; the right axis is the “Steering Control Effort (PWM)” where 90 is steer straight, greater than  90 is turn right, less than 90 is turn left. This team noted in their report that the steering was well tuned with little steady state error. They also mentioned that they used exponential filters to reduce oscillation in the fuzzy controllers. In their report, the team also included plots of throttle control to maintain following distance. 
This same team demonstrated higher order thinking in this way: they observed that fuzzy control required more microcontroller resources. They logged the control loop refresh rate and the program memory required so that PID and Fuzzy methods can be compared on this basis. This is an important consideration and this pair of students addressed it even though it was not required nor mentioned in class.

\subsection{Results of the survey of students about learning}
An end-of-course survey was conducted to determine students' perceptions about the controller comparative study.   A total of 24 students (6 from Fall 2017 and 18 from Fall 2018) responded to the survey.  There were ten statements in the survey:  five designed to assess whether the students felt they exercised skills associated with higher order thinking and five designed to gauge the effectiveness in which the project was structured and administered.  All statements used a Likert scale for agreement with five equally-spaced response options from strongly disagree to strongly agree.  For both semesters, the surveys were conducted either during the last lesson or shortly after the conclusion of the course.  The survey responses are summarized as bar charts.   Each bar represents the average Likert score from the 24 respondents based on a weighted average, and the error bars indicate the standard error of each mean.  The agreement response options in the figures were given numerical weights for the purpose of plotting: Strongly Disagree (SD) = 1, Disagree (D) = 2, Neutral (N) = 3, Agree (A) = 4, and Strongly Agree (SA) = 5.  

\begin{figure}[t]
\centering
\includegraphics[width=3.6in]{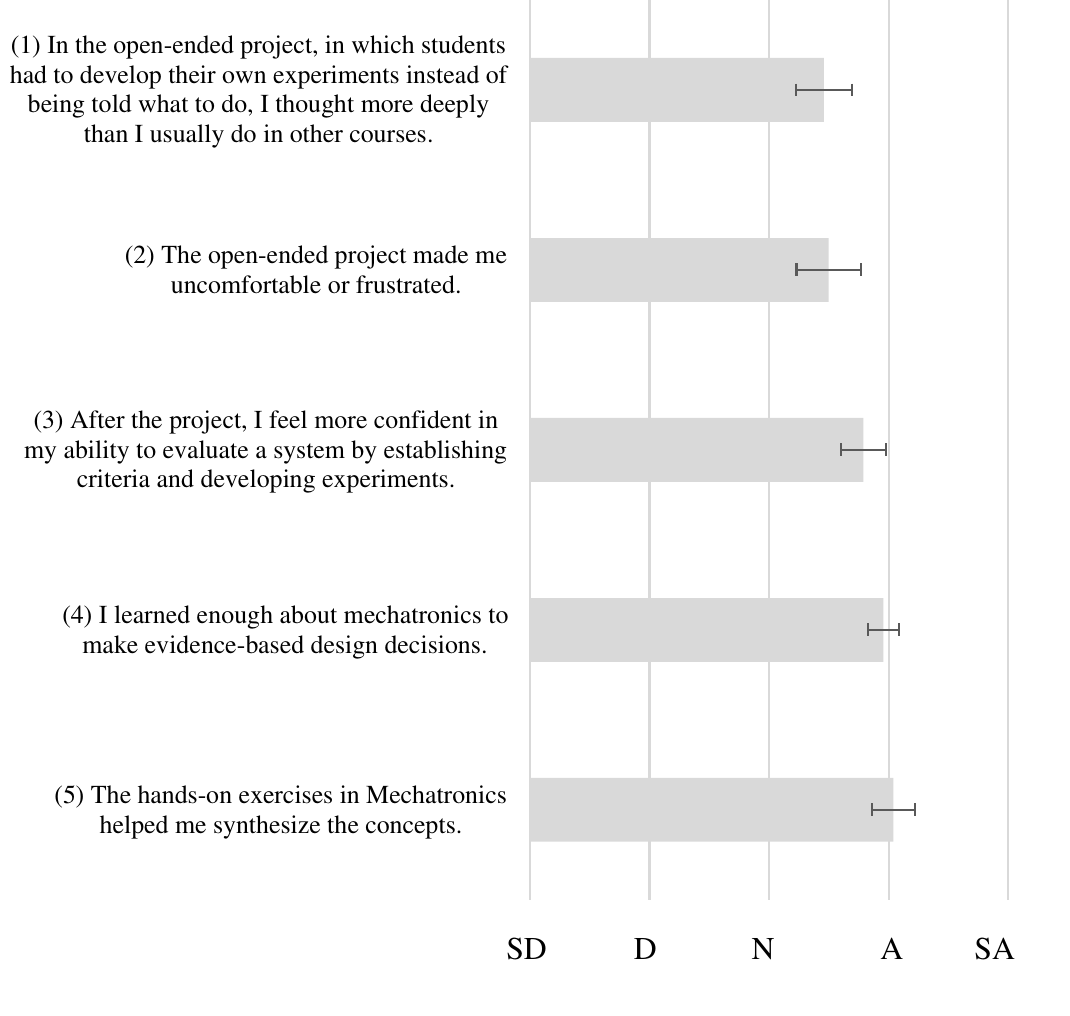}
\caption{Survey statements for assessing the effectiveness of the project in getting the students to practice higher order thinking skills. SD = Strongly Disagree, D = Disagree, N = Neutral, A = Agree, and SA = Strongly Agree.}
\label{effectiveness}
\end{figure}

Fig. \ref{effectiveness} shows the survey responses to a series of statements designed to assess the efficacy of the comparative study in getting the students to reach higher orders of thinking (i.e., use skills associated with analysis, synthesis, and evaluation).  The average responses from the students are ordered in the figure based on an ascending level of agreement.  Nearly all the averaged Likert scores fell between ‘neutral’ and ‘agree’, and the majority tended to be closer to ‘agree’.  When asked whether the students had to think more deeply during the open-ended project than they typically do in other courses, they tended to agree.  Thinking at a higher level can be challenging and somewhat uncomfortable for undergraduate students due to these skills not being practiced frequently at the undergraduate level, and the average response to the second survey statement in Fig. \ref{effectiveness} corroborates this generalization.  The project was designed to get the students out of their ‘comfort zone’ and get them to think more critically about their approach without providing them a step-by-step procedure to evaluate the controllers.  The next two survey responses (3 and 4) in Fig. \ref{effectiveness} confirm that the students tended to agree that the project had them exercise an important skill they will need as engineers:  that is, the ability to frame their own experiments using engineering principles to make design decisions.  Finally, the last statement in Fig. \ref{effectiveness} was formed to assess whether the students felt the hands-on projects throughout the course helped them to interrelate the concepts and to bring together the learning objectives in a coherent fashion; as the average response shows, the majority of the students felt that the series of projects in the course helped them to better synthesize the learning objectives (i.e., reach Bloom’s higher order of thinking through the synthesis or interrelation of concepts).

\begin{figure}[t]
\centering
\includegraphics[width=3.6in]{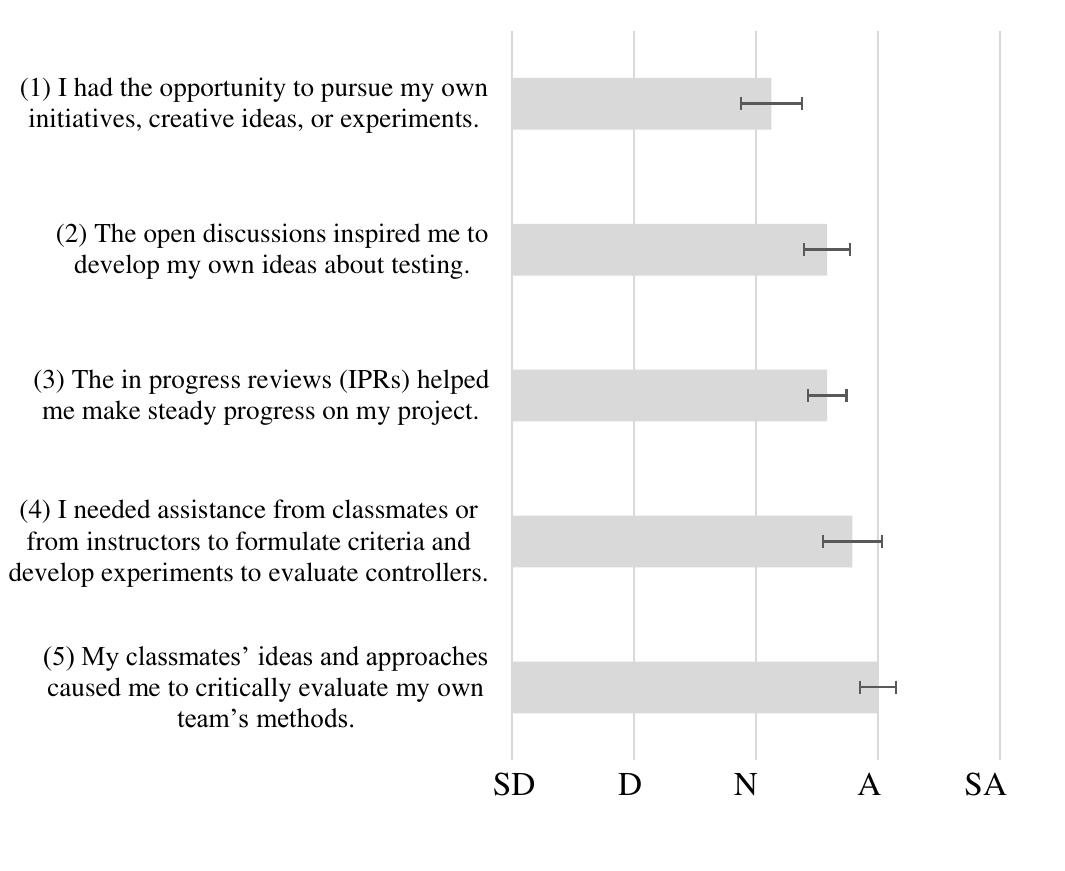}
\caption{Survey statements for assessing the approach of the project such as the in-progress review briefings delivered by the students.}
\label{approach}
\end{figure}

Fig. \ref{approach} shows the survey responses to a series of statements intended to assess our approach in administering the comparative study.  The first response shows that the students were mostly neural about whether they had the creative freedom to pursue their own ideas or set of experiments during the final project.  When the project was administered in the Fall of 2017, the students were instructed to find the best controller, either PID or fuzzy, for automating the steering and velocity of a robot designed to follow a leader.  However, we felt that limiting the project to only the leader-follower application may have been restricting the creative freedom of the students.  As a result, we changed our guidance on the final project in the Fall of 2018 to allow the students the flexibility to propose other applications for their controller comparative study, and proposals required instructor approval to ensure it met the intent of comparing the performance of PID and fuzzy control in a relevant application setting.  Thus, in the future, we expect the students to show more agreement with this survey statement given the added flexibility to choose an application of their liking for conducting a side-by-side comparison of control techniques. 

The remaining statements in Fig. \ref{approach} were designed to assess the effectiveness of the in-progress reviews (IPRs).  Responses to statements 2 and 5 show that the IPRs, which encouraged openness and the exchange of creative ideas, was effective in getting the students to actively listen to their peers’ presentations, ask relevant questions, and reflect on their own approach.  In many cases, groups would modify their planned approach to conduct the evaluation experiences after hearing innovative ideas presented by others during the IPRs.  And to confirm that assessment, the response to statement 4 shows that most of the students needed some form of help from classmates or instructors to fully formulate their evaluation experiments.       

\section{Conclusions}
The organization of a course and the details of what the students must do can be shaped to spur students toward high-level thinking. From the point of view of the authors as engineering educators, ``high-level thinking" means the ability to establish decision criteria, gather evidence, compare alternatives, and make evidence-based decisions. The course is organized with a sequence of lectures and graded requirements that provides students with the knowledge and the structure to develop their own experiments that measure their products against criteria that they themselves establish. The highest level of Bloom’s Taxonomy, \textit{evaluation}, is about making judgments of the relative values of different courses of action. The course provides the students with the opportunity to make a judgment about the relative benefits of two controller designs. Such judgment relies upon a foundation of knowledge, synthesis of an experimental plan with criteria, and analysis of evidence.

Many factors contribute to the success of this course. Its placement in the fall of the senior year allows students to develop the requisite knowledge foundation. Its structure provides the students a learning framework. Its grouping of students of dissimilar majors fosters teamwork and interdisciplinary thinking. Its hands-on nature helps students develop confidence. And finally, its teaching and support staff are dedicated to achievement of the learning objectives.   

The final project---comparison of controllers for an autonomous ``follower" robot---is the most important feature of the course. The project handout indicates no clear ``best" approach to the solution, so students must use their judgment. Timely hands-on exercises throughout the semester help students develop skills need for the project.  Periodic reviews with open discussion keep students engaged with each other and on schedule. Small team size assures accountability. Real world application provides motivation. Students have shown creativity in developing tests to demonstrate how their product meets specifications, or how one controller is better than another. This creativity is a mark of high-level thinking.

\ifCLASSOPTIONcaptionsoff
  \newpage
\fi

\vfill
\pagebreak

\section*{Biographical Information}
\begin{IEEEbiography}%
[{\includegraphics[width=1in,height=1.25in,clip,keepaspectratio]{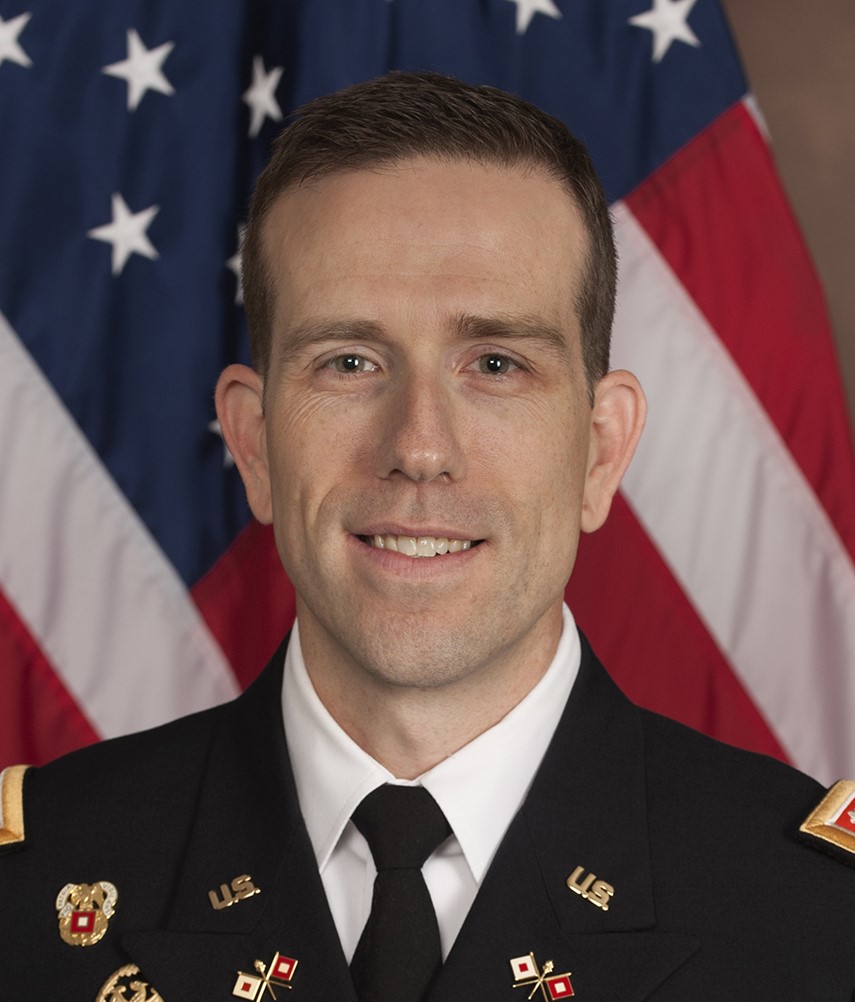}}]%
{LTC Christopher J. Lowrance} is the Chief of Artificial Intelligence (AI) for Autonomous Maneuver at the Army Artificial Intelligence Task Force.  He serves as a technical project manager as well as the Army’s Service Component lead aligned with the Joint Artificial Intelligence Center’s National Mission Initiative on Maneuver and Fires.
Since entering active duty in 2000, he has held multiple command and staff positions as a Signal Officer and Network Systems Engineer.  Most recently, LTC Lowrance served as an Associate Professor in the Department of Electrical Engineering and Computer Science (EECS) and Senior Researcher in the Robotics Research Center (RRC) at the United States Military Academy.  LTC Lowrance's education includes a Bachelor’s Degree in Electrical Engineering (EE) from the Virginia Military Institute (VMI), Master’s Degree in EE from the George Washington University, and Ph.D. from the University of Louisville in Computer Science and Engineering.  He has led multiple research projects related to robotics, machine learning, fuzzy control, and ad hoc networks and has co-authored over 25 peer-reviewed journal and conference publications.  

\end{IEEEbiography}
\begin{IEEEbiography}%
[{\includegraphics[width=1in,height=1.25in,clip,keepaspectratio]{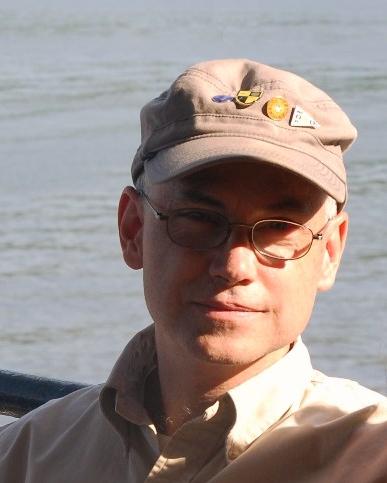}}]%
{Dr. John  R. Rogers} is an Associate Professor in the Department of Civil and Mechanical Engineering at the United States Military Academy, West Point, New York, where he serves the mission to educate train and inspire the nation's young leaders. He teaches courses in design, dynamics, and mechatronics; he mentors student research and he advises design projects. He is an engineer, an educator, a builder, a problem solver, and a broken-thing fixer. Dr. Rogers solves problems using a balance of theory and practical experimentation. He has expertise in mechanical design, electronics, application of sensors and actuators, and microcontrollers. He is interested in simulation of the human musculoskeletal system for exosuit design. Dr. Rogers is a researcher in the USMA Robotics Research center, and a member of the Institutional Review Board for human research subjects protection. He earned the PhD degree in Mechanical Engineering from Rensselaer Polytechnic Institute.
\end{IEEEbiography}
\vfill

\end{document}